\documentclass[oldversion]{aa}

\usepackage{graphicx}

\def\smallsun{\hbox{$_\odot$}}
\def\degr{\hbox{$^\circ$}}
\def\arcmin{\hbox{$^\prime$}}
\def\arcsec{\hbox{$^{\prime\prime}$}}
\def\cm3{cm$^{-3}$}
\def\ebv{$E_{\mathrm{B-V}}=$}

\begin{document}


\title{Abundances of Planetary Nebula NGC\,2392\thanks{Based on
    observations with the Spitzer Space Telescope, which is operated
    by the Jet Propulsion Laboratory, California Institute of
    Technology}}

\author{S.R.\,Pottasch\inst{1} \and J.\,Bernard-Salas\inst{2} \and
  T.L.\,Roellig\inst{3}}

\offprints{pottasch@astro.rug.nl} 

\institute{Kapteyn Astronomical Institute, P.O. Box 800, NL 9700 AV
  Groningen, the Netherlands \and Center for Radiophysics and Space
  Research, Cornell University, Ithaca, NY 14853 \and NASA Ames
  Research Center, MS 245-6, Moffett Field, CA 94035-1000}

\date{Received date /Accepted date}

\abstract{The spectra of the planetary nebula \object{NGC\,2392} is
  reanalysed using spectral measurements made in the mid-infrared with
  the {\em Spitzer Space Telescope}. The aim is to determine the chemical
  composition of this object.  We also make use of {\em IUE} and ground
  based spectra.  Abundances determined from the mid-infrared lines,
  which are insensitive to electron temperature, are used as the basis
  for the determination of the composition, which are found to
  differ somewhat from earlier results. The abundances found,
  especially the low value of helium and oxygen, indicate that the
  central star was originally of rather low mass. Abundances of
  phosphorus, iron, silicon and chlorine have been determined for the
  first time in this nebula. The variation of electron temperature in
  this nebula is very clear reaching quite high values close to the
  center.  The temperature of the central star is discussed in the light
  of the high observed stages of ionization. The
  nebular information indicates the spectrum of the star deviates
  considerably from a blackbody.}


\authorrunning{Pottasch et al.}
\titlerunning{Abundances in NGC\,2392}  

\maketitle

\section{Introduction}

\object{NGC\,2392} (PN~G197.8+17.3) is a bright planetary nebula with
a rather high radial velocity, located in the direction of the
galactic anticenter. A photograph of the nebula is shown in Fig.\,1. The
nebula has a bright inner ellipsoidal, almost round,
shell of about 18\arcsec~diameter. This is surrounded by an almost
spherical region of considerably lower emission with a diameter of
about 40\arcsec. Both the inner and outer regions contain some
structure.  Analysis of the kinematics of the nebula (O'Dell \& Ball,
\cite{odell}) indicates that the inner region is expanding quite
rapidly with respect to the outer region which is also slowly
expanding.  The nebula is located 17 degrees above the galactic plane and has 
only a small extinction. Because of its brightness
it is considered a nearby nebula: most of the uncertain distance
estimates vary between 1 and 2 kpc.

The nebula has a bright central star (V=10.53) which has been
studied by several authors. Pauldrach et al. (\cite{paul}) have
compared the stellar spectrum with a model atmosphere and conclude
that the star has an effective temperature T$_{eff}$=40\,000 K. While
this is slightly higher than the hydrogen Zanstra temperature of about
36\,000 K, it is considerably less than the ionized helium Zanstra
temperature which is close to 70\,000 K. Pauldrach et al.
(\cite{paul}) have noted this discrepancy and admit that they are
unable to explain the additional He-ionizing photons which are needed to
explain the high T$_z$(HeII). Tinkler \& Lamers (\cite{tinkler}) prefer 
assigning a much higher effective temperature to the star based on the high 
ionized helium Zanstra temperature and the high energy balance temperature: 
T$_{eff}$=73\,000 K. It is clear
that the excitation in the nebula is much higher than can be accounted
for by a temperature of T$_{eff}$=40\,000 K. This was already noted by
Natta et al. (\cite{natta}) who showed that higher temperature
ionizing radiation was necessary for all ions found in the nebula: the
higher the ionization potential of the ion studied the higher the
ionizing radiation temperature must be. Why this result is not found
in the analysis of the stellar spectrum is not understood.

The purpose of this paper is to study the element abundances in this
nebula with the help of mid-infrared spectra, in the hope that the
chemical abundances will shed some light on the evolution of this
nebula-central star combination. Abundances have been studied earlier
by Barker (\cite{barker}) and by Henry et al. (\cite{henry}).  Both of
these groups use optical nebular spectra (taken by themselves) and
ultraviolet {\em IUE} spectra. Barker (\cite{barker}) has measurements taken
at five different positions in the nebula. He uses low dispersion {\em IUE}
measurements taken with the small aperture (3\arcsec~diameter).  Henry
et al. (\cite{henry}) also take spectra at five different positions in
the nebula but they use low dispersion {\em IUE} measurements taken with the
large aperture (10\arcsec x 23\arcsec). No abundance changes as
function of the position in the nebula are found by either group. Both
groups agree on the abundances of N, O and Ne to within 20\%. However
their carbon abundances differ by more than a factor of 6 and the
helium abundances differ by almost 30\%. Henry et al. (\cite{henry})
do not discuss the abundance of other elements, while Barker also
determines sulfur and argon.

We have measured the spectrum of NGC\,2392 in the mid-infrared with
the IRS spectrograph of the {\em Spitzer Space Telescope} (Werner et
al.\,\cite{werner}). The use of the mid-infrared spectrum permits a
more accurate determination of the abundances.  The reasons for this
have been discussed in earlier studies (e.g. see Pottasch \& Beintema
\cite{pott1}; Pottasch et al.  \cite{pott2}, \cite{pott4}; Bernard
Salas et al. \cite{bernard}), and can be summarized as follows.

The most important advantage is that the infrared lines originate from
very low energy levels and thus give an abundance which is not
sensitive to the temperature in the nebula, nor to possible
temperature fluctuations. Furthermore, when a line originating from a
high-lying energy level in the same ion is observed, it is possible to
determine an average electron temperature at which the lines in that
particular ion are formed. When this average temperature for many
ions is determined, it is possible to make a plot of electron
temperature against ionization potential, which can be used to
determine the average electron temperature (or range of temperatures) for
which ions of a particular ionization potential are observed. As will
be shown, this is particularly important in NGC\,2392 for which a
large gradient in electron temperature is found. This has not been
taken into account in the earlier abundance studies of this nebula.

Use of the infrared spectra have further advantages. One of them is
that the number of observed ions used in the abundance analysis is
approximately doubled, which removes the need for using large
Ionization Correction Factors (ICFs), thus substantially lowering the
uncertainty in the abundance. A further advantage is that the
extinction in the infrared is almost negligible, eliminating the need
to include sometimes large correction factors. This is not very
important in this nebula. The number of elements which can be measured
increases when including the infrared. In NGC\,2392 phosphorus, iron
and silicon can be measured in the mid-infrared.  This nebula has not
been measured by {\em ISO}, because this part of the sky was not available
to this satellite.

This paper is structured as follows. First the Spitzer spectrum of
NGC\,2392 is presented and discussed (in Sect.\,2). Then the intrinsic
H$\beta$ flux is determined using both the measurements of the
infrared hydrogen lines and the radio continuum flux density
(Sect.\,3). The visible spectrum of the nebula is presented in Sect. 4
together with a new reduction of the ultraviolet ({\em IUE}) spectrum.
This is followed by a discussion of the nebular electron temperature
and density and the chemical composition NGC\,2392 (Sect.\,4). A
comparison of the resultant abundances with those in the literature is
given in Sect.\,5. In Sect.\,6 the central star is discussed
especially in relation to the nebular spectrum.  In Sect.\,7 a general
discussion and concluding remarks are given.

\section{The Infrared spectrum}
\subsection{IRS observations of NGC\,2392}

The observations of NGC\,2392 were made using the Infrared
Spectrograph (IRS, Houck et al.\,\cite{houck}) on board the {\em
  Spitzer Space Telescope} with AORkeys of 4108544 (on target) and
4108800 (background). The data were processed using the s14.0 version
of the pipeline and using a script version of {\em Smart} (Higdon et
al.\,\cite{higdon}). The reduction started from the {\em droop} images
which are equivalent to the {\em bcd} images and just lack stray-cross
removal and flatfield. The tool {\em irsclean} was used to remove
rogue pixels. The different cycles for a given module were combined to
enhance S/N.  At this point the background images were subtracted to
remove the sky contribution. Then the resultant HR images were
extracted using full aperture.  The final spectrum is shown in Fig. 2.

\begin{figure}
  \centering
    \includegraphics[width=8cm,angle=0]{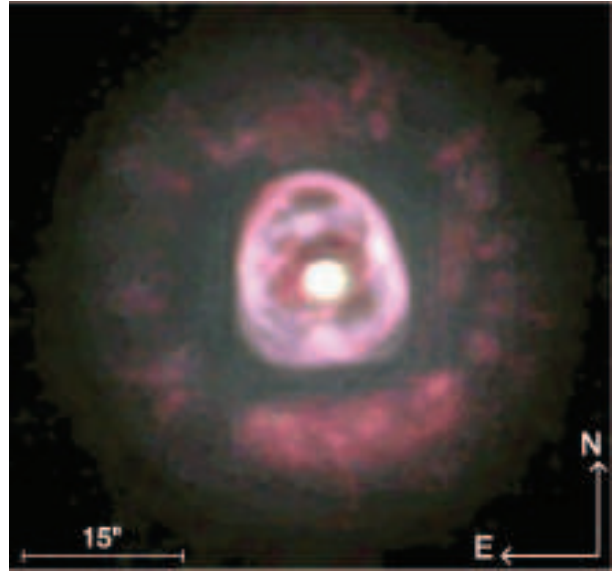}
    \caption{An HST image of NGC\,2392.}
  \label{m142_f}
\end{figure}

The great advantage of the IRS high resolution spectra compared to the
ISO SWS spectra is the very high sensitivity of the IRS. Otherwise the
two instruments are quite comparable. The IRS high resolution spectra
have a spectral resolution of about 600. The IRS high resolution
measures in two spectral ranges with two different modules: the short
high module (SH) going from 9.9$\mu$m to 19.6$\mu$m and the long high
module (LH) from 18.7$\mu$m to 37.2$\mu$m. The SH has a slit size of
4.7\arcsec x 11.3\arcsec, while the LH is 11.1\arcsec x 22.3\arcsec.

\setcounter{figure}{1}

\begin{figure*}
  \centering
    \includegraphics[width=12cm,angle=90]{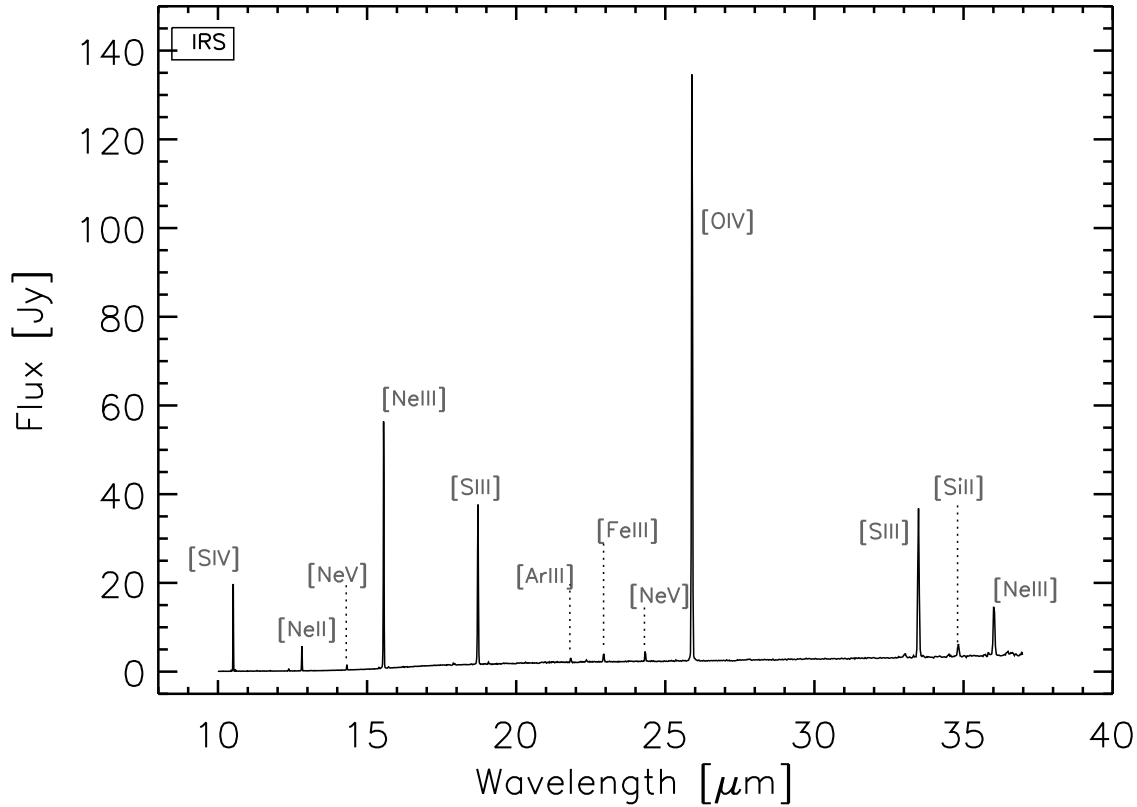}
    \caption{Observed Spitzer-IRS high-resolution spectrum of NGC\,2392.
      The most prominent lines are labeled in the figure.}
  \label{irs_f}
\end{figure*}

The IRS measurement of NGC\,2392 was centered at RA(2000)
07$^{h}$29$^{m}$10.77$^{s}$ and Dec(2000)
+20\degr54\arcmin42.5\arcsec. This is almost the same as the value
measured by Kerber et al. (\cite{kerber}) of RA(2000)
07$^{h}$29$^{m}$10.768$^{s}$ and Dec(2000)
$-$28\degr54\arcmin42.44\arcsec. Thus the IRS measurement was well
centered on the nebula and both the LH and SH diaphragms measured
essentially the inner, brighter nebula. Since the LH diaphragm is
larger, more of the nebula is seen in this diaphragm, therefore a
correction must be made to bring the two measurements to the same
scale. This was done by making use of the fact that the two
spectrographs had a small wavelength region in common at about
19$\mu$m. To make the continuum emission at this wavelength equal, the
SH emission had to be increased by 4.3. This is close to the
difference in area between the two slits (4.7).  The measured emission
line intensities are given in Table 1, after correcting the SH
measurements by the factor 4.3 in the column labeled 'intensity'.  The
fluxes were measured using the Gaussian line-fitting routine and
derived for each {\em nod} position independently. The uncertainty in
the fluxes was assumed to be the largest of either the difference
between the flux in the {\em nod} positions, or the uncertainty in the
fit.  The last column gives the ratio of the intensity to H$\beta$
where the H$\beta$ is found from the strongest hydrogen line(s)
measured in the IRS spectrum. It has been assumed that the ratio of
the sum of the two hydrogen lines at 12.372$\mu$m to H$\beta$ has a
value of 8.80 x 10$^{-3}$, which is the value given by Hummer \&
Storey (\cite{hummer}) for an electron temperature of 15\,000 K. No
correction for extinction is made since it is very small. This means
that the value of H$\beta$ through the LH diaphragm is 2.87x10$^{-11}$
erg cm$^{-2}$s$^{-1}$ and the correction for reddening has been
included. Note that the three lines with wavelength less than 10$\mu$m
are measured in the low resolution spectrograph.

\begin{table}
  \caption[]{IRS spectrum of \object{NGC\,2392}. The measured line intensity is
    given in Col.3. The last column gives the ratio of the line
    intensity to H$\beta$(=100).}

\begin{center}
\begin{tabular}{lrcc}
\hline
\hline
Identification &  $\lambda$($\mu$m) & Intensity$^{\dagger}$ & I/H$\beta$ \\
\hline

\ion{[Ar}{ii]} & 6.989 & 18.3$\pm$3.5 & 0.64 \\
\ion{H}{i}(6-5) & 7.481 & 47.5$\pm$2.2 & 1.66 \\
\ion{[Ar}{iii]} & 8.995 & 231$\pm$8.8 & 8.05 \\
\ion{[S}{iv]} & 10.511 & 950$\pm$56  & 33.0\\
\ion{H}{i}(9-7) & 11.309 & 8.12$\pm$0.47 & 0.28 \\
\ion{[Cl}{iv]} & 11.762 & 4.69$\pm$0.60 & 0.163 \\
\ion{H}{i} (7-6+11-8) & 12.374 & 25.3$\pm$1.4 &    \\
\ion{[Ne}{ii]} & 12.815 & 220$\pm$10  & 8.63 \\\
\ion{[Ar}{v]} & 13.096 & $<$2.6 &     \\
\ion{[Ne}{v]} & 14.323 & 44.6$\pm$0.86 & 1.56 \\
\ion{[Ne}{iii]} & 15.556 & 1930$\pm$30.8 & 67.2 \\
\ion{H}{i} 10-8 & 16.219 & 4.2$\pm$2.0 & 0.146: \\
\ion{[P}{iii]}  & 17.893 & 16$\pm$20 & 0.56: \\
\ion{[Fe}{ii]}  & 17.943 & 10.0$\pm$40 & 0.35: \\
\ion{[S}{iii]} & 18.718 & 1098$\pm$27  & 38.2 \\
\ion{[Cl}{iv]} & 20.319 & 4.91$\pm$0.36  & 0.167 \\
\ion{[Ar}{iii]} & 21.827 & 25.6$\pm$1.96 & 0.89 \\
\ion{[Fe}{iii]} & 22.934 & 44.7$\pm$1.37 & 1.55 \\
\ion{[Ne}{v]} & 24.326 & 44.5$\pm$3.25 & 1.55 \\
\ion{[O}{iv]} & 25.895 & 2634$\pm$119  & 88.4 \\
\ion{H}{i} 9-8 & 27.806 & 6.76$\pm$1.69 & 0.236 \\
\ion{[Fe}{iii]} & 33.043 & 14.7$\pm$1.31 & 0.511 \\
\ion{[S}{iii]} & 33.487 & 530$\pm$11.5  & 18.4 \\
\ion{[Si}{ii]} & 34.825 & 45.4$\pm$1.69 & 1.58 \\
\ion{[Ne}{iii]} & 35.941 & 97.7$\pm$115: & 3.42: \\

\hline

\end{tabular}
\end{center}

$^{\dagger}$ Intensities measured in units of 10$^{-14}$
erg~cm$^{-2}$~s$^{-1}$.  The intensities below 19$\mu$m have been
increased by a factor of 4.3 to bring them on the same scale as the LH
intensities measured through a larger diaphragm. The intensities below 10$\mu$m
are measured with the low resolution instrument.
 
: Indicates an uncertain value.
 
\end{table}

There are other ways of determining the factor which accounts for the
emission missing in the SH diaphragm. The method given above, that of
making use of the fact that the continua in the region of wavelength
overlap should be equal, is theoretically satisfying. This is because the
continuum is due primarily to dust which probably has the same properties
throughout the nebula. But to check this any pair of
lines can be used as long as they originate from the same ion and two
other conditions exist. One is that each of the lines is measured in a
different diaphram. The other is that the theoretical value of the
ratio of the two lines is only weakly dependent on the electron
temperature and density. Five candidate pairs of lines exist. One is
the ratio of the \ion{[Ne}{iii]} lines at $\lambda$15.5 and
$\lambda$36.0$\mu$m. This cannot be used in this case because the
$\lambda$36.0$\mu$m line is very uncertain.  The \ion{[Ne}{v]} and
\ion{[Cl}{iv]} ions both have a pair of lines, one of which is in the SH
range and one of which is in the LH range. The predicted line ratio is
only slightly dependent on the nebular properties so that they may be
used; they are consistent with the value of 4.3. The hydrogen
lines are rather weak but they also are consistent with a value of 4.3
with an error of about 20\%. It thus appears that the total emission
in the SH wavelength band is obtained by increasing the measured
intensities by a factor of 4.3$\pm$0.2. This is quite close to the ratio of 
the areas of the two diaphragms which is 4.6.

\section{Extinction}

There are several methods for obtaining the extinction: (1) comparison
of radio emission with H$\beta$ flux, (2) comparison of observed and
theoretical Balmer decrement, (3) dip at $\lambda$2200\AA, (4)
photometry of the exciting star. All of these methods are applicable
here. First we will discuss the radio emission and how the H$\beta$
flux is obtained.

\subsection{The 6\,cm radio emission}

The 6\,cm flux density has been measured by Becker et al.
(\cite{becker}) and Gregory \& Condon (\cite{greg}).  Becker et al.
(\cite{becker}) find a value of 242 mJy and Gregory \& Condon
(\cite{greg}) give 246 mJy. The nebula has also been measured at
21\,cm by Condon \& Kaplan (\cite{condon}) who find a value of 280.9
mJy, which implies a 6\,cm flux density of 238 mJy if the nebula
is optically thin. We will use a value of 242 mJy. Using values of
T$_e$ and helium abundance determined below together with the equation
quoted in Pottasch (\cite{potta}) this implies an H$\beta$ flux of
6.83x10$^{-11}$ erg~cm$^{-2}$~s$^{-1}$, indicating that about 42\% of
the nebula is being measured in the LH slit.

\subsection{Extinction}

The extinction determined from the Balmer decrement differs somewhat
according to the author. Henry et al. (\cite{henry}) give C=0, Aller
\& Czyzak (\cite{aller}) give C=0.15, Barker (\cite{barker}) finds
C=0.12.

The photometry of the star gives a higher extinction. Using B=10.38
and V=10.53, B$-$V=$-$0.15. The intrinsic value of B-V for a hot star
is $-$0.32 giving a value of \ebv0.17 or C=0.25. The extinction found
from the $\lambda$2200\AA~dip is \ebv0.08 or C=0.12 (Pottasch et al.
\cite{pott5}).

The measured value of the integrated H$\beta$ flux is 4.07x10$^{-11}$
erg~cm$^{-2}$~s$^{-1}$ (Cahn et al. \cite{cahn}). Using the value of
H$\beta$ from the radio measurements leads to a value of C=0.225 or
$E_{\mathrm{B-V}}$=0.154. This is slightly higher than the values
found from the Balmer decrement and the $\lambda$2200\AA~dip but it
agrees with the extinction found from the stellar photometry. We find 
that the radio/H$\beta$ method is the most accurate determination, 
so that this value will be used when necessary in this paper. But the 
extinction is so low that it does not affect any results in this paper.

\subsection{The visual spectrum}

The visual spectrum has been measured by several authors. The highest
resolution spectrum is by Henry et al. (\cite{henry}), and reliable
spectra have also been reported by Barker (\cite{barker}) and Aller \&
Czyzak (\cite{aller}). The measurements of Henry et al. (\cite{henry})
and those of Barker (\cite{barker}) are made at several different
places in the nebula but there do not appear to be important
differences between them. The results are shown in Table 2 where Cols.
3, 4 and 5 give the intensities measured by the various authors
relative to H$\beta$=100 for those lines which are of interest. The
line intensities have been corrected by the individual authors for
extinction. Henry et al.(\cite{henry}) felt that no extinction
correction was necessary, the other authors using the reddening curve
of Seaton (\cite{seaton}) find the extinctions which are listed in the
previous section. No attempt has been made to use a common extinction
correction because all give a correct Balmer decrement. In the last
column average values are given; more weight is given to the
measurements of the first two authors. All authors estimate that the
strongest lines have a 10\% error, the intermediate strength lines
(about 5\% of H$\beta$) have about 20\% error and the weakest lines
have about 30\% error. The average intensities have about the same
error.
     
\begin{table}[h]
\caption[]{Visual Spectrum of NGC\,2392.}
\begin{center}
\begin{tabular}{llcccc}
\hline
\hline
\multicolumn{1}{c}{$\lambda$} & Ion & \multicolumn{3}{c}{Intensities$^{\dagger}$}& Average\\ \cline{3-5}
\multicolumn{1}{c}{(\AA)}& & (1) & (2)& (3) & Intens.\\
\hline

3727$^{\ast}$   & \ion{[O}{ii]} & 107 & 120  & 108  & 110 \\
3869 & \ion{[Ne}{iii]} & 100 & 130 & 90  & 105 \\
4101 & \ion{H$\delta$}& 24.0 & 26.5 & 25.2 & 25.2 \\
4267 & \ion{C}{ii}    &    &  0.3: &   & 0.3: \\
4340 & \ion{H$\gamma$}& 46.0 & 48.0 & 48.0 & 47.2 \\
4363 & \ion{[O}{iii]} & 14 & 25   & 19.5  & 19.5 \\
4686 & \ion{He}{ii}  & 38 & 38 & 35 & 37 \\
4740 & \ion{[Ar}{iv]} &      &  2.2 & 2.5 & 2.3 \\
4861 & \ion{H$\beta$} & 100 & 100 & 100 & 100\\
5007 & \ion{[O}{iii]} & 950 & 1200 & 1260 & 1150\\
5517 & \ion{[Cl}{iii]} & 0.6 & 0.72  & 0.65   &  0.65 \\
5538 & \ion{[Cl}{iii]} &       & 0.68  & 0.50   & 0.59 \\
5755 & \ion{[N}{ii]}  & 2.0  & 1.6 & 1.51  & 1.6 \\
5876 & \ion{He}{i}    & 6.8   & 7.8 & 7.45  & 7.4 \\
6312 & \ion{[S}{iii]} & 2.5  & 3.6  & 3.3  & 3.2 \\
6563 & \ion{H$\alpha$}& 275  & 295  & 283  & 285 \\
6584 & \ion{[N}{ii]}  & 95  & 95  & 85.5   & 92 \\
6717 & \ion{[S}{ii]}  & 8.0 & 6.7  & 4.8  & 6.7 \\
6731 & \ion{[S}{ii]}  & 10.0 & 7.9  & 7.8  & 8.6 \\
7005 & \ion{[Ar}{v]}   &     &      &  0.15 & 0.15 \\
7135 & \ion{[Ar}{iii]}& 14 & 14.3  & 12.3  & 14 \\
7237 & \ion{[Ar}{iv]}  &     &   & 0.19:  & 0.19: \\
8045 & \ion{[Cl}{iv]}  & 0.4    &   & 0.47  & 0.45 \\
9532 & \ion{[S}{iii]}  &  91  &    &      &  91 \\

\hline
\end{tabular}
\end{center}

$^{\dagger}$ References; (1) Henry et al. (\cite{henry}), (2) Barker (\cite{barker}), (3) Aller \& Czyzak (\cite{aller}).\\ 
(:) indicates uncertain values.\\ 
$^{\ast}$ This is a blend of $\lambda$3726 and $\lambda$3729 lines.
\end{table}

\subsection{The IUE ultraviolet spectrum }

There are 80 low resolution {\em IUE} observations of this nebula as well as
six high resolution observations. Most were taken with the large
aperature (10\arcsec x 23\arcsec) with varying exposure times, but
quite a few were taken with a small aperature (3\arcsec~diameter). The
small aperature measurements were used by Barker (\cite{barker}), who
did not find any important abundance variations in the nebula. We find
the small aperature measurements are too noisy and we do not use them.
The large aperture measurements do not cover the entire nebula. Most
of the measurements which exclude the central star cover about 10\% to
15\% of the nebula. We have used three of the best observations made
with long exposure times. These are SWP05231 and SWP25236 for the
short wavelength region and LWR04515 for the longwavelength region. We
have checked that the strongest lines are not saturated by measuring
these same lines on spectra with shorter exposure times. They are
shown in Table 3. The uncertainties are about the same as given by Henry et al.
(\cite{henry}): about 10\% for the strongest lines and about 25\% for the 
weaker lines.

The extinction correction was made by assuming a theoretical ratio for
the \ion{He}{ii} line ratio $\lambda$1640/$\lambda$4686\,\AA~ at
T=12\,500 K and an N$_e$ of 10$^3$ cm$^{-3}$. The ratio of $\lambda$1640 to 
H$\beta$ can then be found using the $\lambda$4686/H$\beta$ ratio in Table 2
which leads to a value of the
$\lambda$1640\,\AA~ line as given in column (2) of Table 3. A further
correction for extinction relative to $\lambda$1640\AA~is then made
using the reddening curve of Fluks et al.(\cite{fluks}) but because of
the small extinction this is never more than 20\%, which indicates the
uncertainties of the UV intensities above the errors of measurement given
above. The results are shown in the last two columns of Table 3.

\begin{table}[htbp]
\caption[]{IUE Spectrum of NGC\,2392.}  
\begin{center}
\begin{tabular}{llccc}
\hline
\hline
\multicolumn{1}{c}{$\lambda$} & Ion &\multicolumn{3}{c}{Intensities}\\
\cline{3-5}
\multicolumn{1}{c}{(\AA)}& & (1) & (2) &  (I/H$\beta$)   \\
\hline

1335 & \ion{C}{ii}   & 10.2  & 11.0  & 16.0 \\
1400 & \ion{O}{iv}  &  36.3  & 36.6 & 53.4 \\
1485 & \ion{N}{iv]} &  24.8  & 24.2  & 35.2 \\
1548 & \ion{C}{iv}  &  77.7  & 75.5  & 110  \\
1640 & \ion{He}{ii}  & 185   & 176  & 256  \\
1663 & \ion{O}{iii]} & 53.8   & 34.6 & 50.4 \\
1750 & \ion{N}{iii]} & 49.4  &  45.4 & 66.1 \\
1881 & \ion{Si}{iii]} & 14.7  &  14.0  & 20.4 \\
1890 & \ion{Si}{iii]} & 12.0  &  11.4  & 16.6 \\
1909 & \ion{C}{iii]} & 156    & 148  & 216  \\
2325 & \ion{C}{ii]}  & 36.4   & 31.8 & 46.4 \\
2423 & \ion{[Ne}{iv]} & 187   & 145  & 211  \\
2472 & \ion{[O}{ii]}  &  20   & 14.7 & 21.5 \\
2512 & \ion{He}{ii}  &  24.2  & 17.3 & 25.2 \\

\hline 
\end{tabular}
\end{center}
(1)Measured intensity from low resolution spectra SWP05231 and LWR04515 in 
units of 10$^{-13}$ erg cm$^{-2}$ s$^{-1}$. \\
(2)Intensity corrected for diaphragm size and extinction in units of
10$^{-12}$ erg cm$^{-2}$ s$^{-1}$. 
I/H$\beta$ is normalized to H$\beta$=100.\\ 
\end{table}

\section{Chemical composition of the nebulae}

The method of analysis is the same as used in the papers cited in the
introduction. First the electron density and temperature as function
of the ionization potential are determined. Then the ionic abundances
are determined, using density and temperature appropriate for the ion
under consideration. Then the element abundances are found for those
elements in which a sufficient number of ionic abundances have been
derived.

\subsection{Electron density}  

The ions used to determine $N_{\mathrm{e}}$ are listed in the first
column of Table\,4. The ionization potential required to reach this
stage of ionization, and the wavelengths of the lines used, are given
in Cols.\,2 and 3 of the table. Note that the wavelength units are
\AA~when 4 ciphers are given and microns when 3 ciphers are shown. The
observed ratio of the lines is given in the fourth column; the
corresponding $N_{\mathrm{e}}$ is given in the fifth column. The
temperature used is discussed in the following section, but is
unimportant since these line ratios are essentially determined by the
density. No density is given from the \ion{Ne}{iii} lines because of
the uncertainty of the 36$\mu$m line intensity. The density from the
\ion{C}{iii]} lines is uncertain because it is possible the stronger of
the two lines may be saturated.

There is no indication that the electron density varies with
ionization potential in a systematic way. The electron density appears
to be about 1200 cm$^{-3}$. The error is about 20\%. It is interesting
to compare this value of the density with the rms density found from
the H$\beta$ line. This depends on the distance of the nebula which
isn't accurately known, and on the angular size of the nebula. For
this calculation we shall use a distance of 1.5~kpc and a diameter of
18\arcsec~for the inner region. This gives an rms density of 2000
cm$^{-3}$ for the inner region. The comparison is difficult because the
measurement may include some H$\beta$ flux from the outer region which
will lower the density. This uncertainty is illustrated by two
different \ion{O}{ii} measurements. Kingsburgh \& Barlow
(\cite{kings}) measured a 3726/3729\AA=1.35 at the center of the
nebula, corresponding to a density of 1650 cm$^{-3}$, while O'Dell \&
Castaneda (\cite{odellc}) measuring at the (brighter) rim of the
nebula, find a value of 1.78 for the ratio, corresponding to a density
of 3000 cm$^{-3}$. We will use a density of 1500 cm$^{-3}$ in further
discussion of the abundances, but any value between 1000 cm$^{-3}$ and
3000 cm$^{-3}$ will give the same values of abundance.

\begin{table}[t]
\caption[]{ Electron density indicators in NGC\,2392.}
\begin{center}
\begin{tabular}{lcccc}
\hline
\hline
Ion &Ioniz. & Lines& Observed &N$_{\mathrm{e}}$ \\
&Pot.(eV) & Used  & Ratio & (cm$^{-3}$)\\
\hline
\ion{[S}{ii]} & 10.4 & 6731/6716 & 1.28  & 1600\\
\ion{[O}{ii]} & 13.6 & 3626/3729 & 1.35 & 1650\\
\ion{[S}{iii]} & 23.3 & 33.5/18.7 & 0.48 & 2600 \\
\ion{[Cl}{iii]} & 23.8 & 5538/5518 & 0.86 & 1200\\
\ion{C}{iii]} & 24.4 & 1906/1909 & 1.46 & 1000:\\ 
\ion{[Ne}{iv]} & 63.5 & 2425/2422 & 1.4 & 1000: \\
\ion{[Ne}{v]} & 97.1  & 24.3/14.3  & 0.99 & 1000: \\

\hline
\end{tabular}
\end{center}
: Indicates uncertain values.

\end{table}

\subsection{Electron temperature}

A number of ions have lines originating from energy levels far enough
apart that their ratio is sensitive to the electron temperature. These
are listed in Table 5 which is arranged similarly to the previous
table. The values are also shown in Fig.\,3 where it can be seen that the 
electron temperature has a strong gradient as a function of
ionization potential. Other PNe have also shown a gradient of electron
temperature but it is especially strong in this nebula, going from a
value of about 10\,000~K for the lowest ionization potentials
(presumably formed in the outer regions of the nebula) through a value
of 14\,000 to 15\,000~K at an ionization potential of about 35~eV, to
a value well over 20\,000~K at ionization potentials above 50~eV. For
these latter ions it is impossible to specify the temperature
accurately, so that the abundance of these ions can only be determined
from the infrared lines since the population of the lower levels is
only slightly dependent on the temperature.


\begin{table}[t]
\caption[]{Electron temperature indicators in NGC\,2393.}
\begin{center}
\begin{tabular}{lcccc}
\hline
\hline
 Ion & Ioniz. & Lines& Observed & $T_{\mathrm{e}}$\\
 & Pot.(eV)& Used &Ratio  & (K) \\
\hline

\ion{[N}{ii]}  & 14.5 & 5755/6584 & 0.017 & 10\,000 \\
\ion{[S}{iii]} & 23.3 & 6312/18.7 & 0.0.084 & 12\,000 \\
\ion{[Ar}{iii]} & 27.6 & 7136/21.8 & 15.8 & 10\,000\\
\ion{[O}{iii]} & 35.1 & 4363/5007 & 0.0165 & 14\,700\\
\ion{[O}{iii]} & 35.1 & 1663/5007  & 0.0496 & 15\,200\\
\ion{[Ne}{iii]} & 41.0 & 3869/15.5 & 1.56 & 13\,500\\
\ion{[O}{iv]}   & 54.9 & 1400/25.9 & 0.44 & 22\,000:\\
\ion{[Ne}{v]}  & 97.1  & 3425/24.3 & 2.2: & 25\,000 \\

\hline
\end{tabular}
\end{center}
: Indicates uncertain value.\\ When the wavelength has 4 ciphers it has the 
units of Angstrom and 3 ciphers is micron.
\end{table}

\begin{figure}
  \centering
    \includegraphics[width=6cm,angle=90]{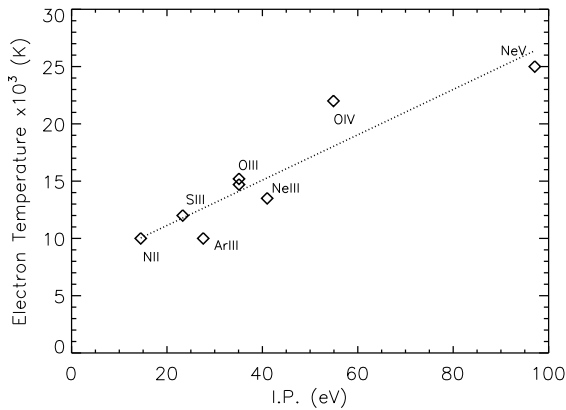}
    \caption{Electron temperature is plotted as a function of the Ionization 
Potential.}
  \label{m142_f}
\end{figure}

\subsection{Ionic and element abundances}

The ionic abundances have been determined using the following
equation:

\begin{equation}
\frac{N_{\mathrm{ion}}}{N_{\mathrm{p}}}= \frac{I_{\mathrm{ion}}}{I_{\mathrm{H_{\beta}}
}} N_{\mathrm{e}}
\frac{\lambda_{\mathrm{ul}}}{\lambda_{\mathrm{H_{\beta}}}} \frac{\alpha_{\mathrm{H_{\beta}}}}{A_{\mathrm{ul}}}
\left( \frac{N_{\mathrm{u}}}{N_{\mathrm{ion}}} \right)^{-1} 
\label{eq_abun}
\end{equation}

where $I_{\mathrm{ion}}$/$I_{\mathrm{H_{\beta}}}$ is the measured
intensity of the ionic line compared to H$\beta$, $N_{\mathrm{p}}$ is
the density of ionized hydrogen, $\lambda_{\mathrm{ul}}$ is the
wavelength of the line, $\lambda_{\mathrm{H_\beta}}$ is the wavelength
of H$\beta$, ${\alpha_{\mathrm{H_\beta}}}$ is the effective
recombination coefficient for H$\beta$, $A_{\mathrm{ul}}$ is the
Einstein spontaneous transition rate for the line, and
$N_{\mathrm{u}}$/$N_{\mathrm{ion}}$ is the ratio of the population of
the level from which the line originates to the total population of
the ion. This ratio has been determined using a five level atom. The atomic
data used is given in the paper of Pottasch \& Beintema (\cite{pott1}. The 
use of a 5 level atom in the case of iron is justified because
collisional rates to higher levels followed by cascade through the lower levels
are quite small in comparison to direct collisions to the lower levels.

The results are given in Table 6, where the first column lists the ion
concerned, the second column the line used for the abundance
determination, and the third column gives the intensity of the line
used relative to H$\beta$=100. The fourth column gives the value of
the electron temperature used for the abundance determination of the
ion, which is shown in the fifth column. This temperature is
determined by plotting the temperatures given in Table 5 as function
of ionization potential, drawing a line through
these points as shown in Fig.\,3, and then reading the temperature of the ion 
under consideration from its ionization potential. The fifth column gives
the ionic abundance assuming the ion is formed at this temperature,
while the sixth column gives the Ionization Correction Factor (ICF),
which has been determined empirically. Notice that the ICF is unity
(or almost unity) for all elements listed in the table except for Fe,
Si, Cl and P.

The error of measurement of the IRS intensities as can be seen in Table 1 is 
usually small, often not more than 5-7\%. In the few cases when the error is 
large this has either been indicated with a ':' or by not using the line.
The correction for adjusting the SH to LH intensity scales and the 
diaphragm size is probably small, about 10\%. This includes the 
assumption that the unmeasured parts of
the nebula have the same spectrum as the measured parts. This has been checked
in the optical region by Barker\,(\cite{barker}) and can be seen to be a
reasonable approximation in the ultraviolet region by comparing the {\em IUE} 
spectra made in different parts of the nebula. The uncertainty of the 
collisional strengths introduces
an error of 10-15\% so that the total error for the ions of neon, sulfur and
argon determined with the IRS measurements is less than 20\%. This will also
be true of the abundances of these elements because the ICF for these elements
is close to unity. The error for the nitrogen and oxygen abundances is 
somewhat higher because the visual and ultraviolet measurements are less
certain. In addition the temperature is more important for these ions and the
total errors may be twice as large. The element abundances are given in the 
last column,  The carbon recombination line abundance is given at
the end of the table. These abundances will be discussed in the next
sections.

The helium abundance has been derived using the theoretical work of
Benjamin et al. (\cite{benjamin}). For recombination of singly ionized
helium, most weight is given to the $\lambda$ 5875\AA~line, because
the theoretical determination of this line is the most reliable.

\begin{table}[h]
  \caption[]{Ionic concentrations and chemical abundances in NGC\,2392.
    Wavelength in Angstrom for all values of $\lambda$ above 1000, otherwise
    in $\mu$m.}


\begin{tabular}{lcccccc}
\hline
\hline
Ion & $\lambda$ &  I/H$\beta$ & T$_e$ & $N_{\mathrm{ion}}$/$N_{\mathrm{p}}$ 
& ICF & $N_{\mathrm{el.}}$/$N_{\mathrm{p}}$\\
\hline

He$^{+}$  & 5875 &  7.4  & 12\,000 & 0.048    &     &  \\
He$^{++}$ & 4686 & 37    & 14\,000 & 0.032    & 1.0 & 0.080 \\
C$^{+}$   & 2324 & 41.7  & 10\,000 & 7.5(-5)  &     &    \\
C$^{++}$  & 1909 & 240   & 11\,500 & 2.3(-4) &     &   \\
C$^{+3}$  & 1548 & 115   & 17\,000 & 1.7(-5)  & 1.0 & 3.3(-4) \\
N$^{+}$   & 6584 & 92    & 10\,000 & 1.8(-5) &     &   \\
N$^{++}$  & 1750 & 62.3  & 12\,500 & 1.5(-4) &     &    \\
N$^{+3}$  & 1485 & 30    & 17\,000 & 1.4(-5) & 1.0 & 1.85(-4) \\
O$^{+}$   & 3727 & 110   & 10\,000 & 5.0(-5) &     &     \\
O$^{++}$  & 5007 & 1150  & 14\,000 & 1.9(-4) &     &    \\
O$^{++}$  & 1663 & 57    & 14\,000 & 2.3(-4)  &     &     \\
O$^{+3}$  & 25.8 & 88.4  & 19\,500 & 2.3(-5)  & 1.0 & 2.9(-4) \\
Ne$^{+}$  & 12.8 & 8.63  & 11\,000 & 1.2(-5) &     &       \\
Ne$^{++}$ & 15.5 & 67.2  & 15\,000 & 4.7(-5) &     &       \\
Ne$^{++}$ & 3869 & 105   & 15\,000 & 3.5(-5) &     &    \\
Ne$^{+3}$ & 2423 & 211   & 21\,000 & 2.4(-5) &     &    \\
Ne$^{+4}$ & 24.3 & 1.55  & 27\,000 & 1.9(-7) & 1.0 & 8.5(-4)  \\  
S$^{+}$   & 6731 & 8.6   & 9\,000  & 4.9(-7)  &     &      \\
S$^{++}$  & 18.7 & 38.2  & 11\,300 & 3.5(-6) &     & \\
S$^{++}$  & 6312 & 3.2   & 11\,300 & 5.2(-6)  &     &     \\
S$^{+3}$  & 10.5 & 33.0  & 14\,000 & 0.87(-6) & 1.0 & 5.0(-6) \\
Ar$^{++}$ & 21.8 & 0.89  & 12\,000 & 1.3(-6) &     &       \\
Ar$^{++}$ & 7135 & 14.0  & 12\,000 & 1.0(-6) &     &         \\
Ar$^{+3}$ & 4740 & 2.3   & 15\,300 & 2.9(-7) &     &       \\
Ar$^{+3}$ & 7237 & 0.19  & 15\,300 & 8.1(-7):  &     &       \\
Ar$^{+4}$ & 7005 & 0.15  & 20\,500 & 1.4(-8)  & 1.2 & 2.2(-6) \\
Cl$^{++}$ & 5538 & 0.59  & 11\,400 & 7.2(-8)  &     &       \\ 
Cl$^{+3}$ & 20.3 & 0.167 & 14\,800 & 0.88(-8) &     &      \\
Cl$^{+3}$ & 8045 & 0.45  & 14\,800 & 2.7(-8) & 1.3 & 1.3(-7) \\
Fe$^{++}$ & 22.9 & 1.55  & 10\,500 & 4.7(-7) &     &      : \\ 
Fe$^{++}$ & 33.0 & 0.51  & 10\,500 & 5.8 (-7) & 1.6 & 8.0(-7) \\
Si$^{+}$  & 34.8 & 1.58  & 9\,000  & 3.8(-7) &     &      \\ 
Si$^{++}$ & 1888 & 37.0  & 10\,400 & 1.4(-5) & 1.3 & 1.9(-5) \\
P$^{++}$  & 17.9 & 0.56  & 10\,500 & 3.4(-8): & 1.9:& 6.5(-8):  \\
C$^{++}$  & 4267 & 0.3   & 11\,500 & 3.4(-4):  &     &  \\
 
\hline
\end{tabular}

Intensities given with respect to H$\beta$=100.

: Indicates uncertain value.

\end{table}

\section{Comparison with other abundance determinations}

Table 7 shows a comparison of our abundances with the most important
determinations in the past 20 years. Only about half of the elements
have been reported before. Good agreement is found for oxygen; this is
because the same electron temperature is used for the most important
oxygen ion. For the other elements the agreement is less good. We find
a rather higher carbon abundance, much higher than found by Barker
(\cite{barker}) and slightly higher than found by Henry et al.
(\cite{henry}). A C/O ratio greater than unity is found but the error
is large enough that it could be slightly less than unity. Nitrogen is
somewhat higher than found by the other authors and clearly a factor
of two higher than the solar abundance. Most of the other elements are
consistent with a galactocentric abundance gradient (-0.085 dex/kpc)
as described for example by Pottasch \& Bernard-Salas (\cite{pbs}).
Iron is depleted by more than a factor of 30 with respect to the solar
abundance, as it is in most of the nebulae already studied. Phosphorus
seems rather strongly depleted as well, but it is uncertain because of
the large measurement error. The comparison made in the table with the
solar abundance is taken from Asplund et al. (\cite{asplund}).  Note
that for sulfur and chlorine more weight has been given to the
abundance determination in meteorites since this determination is more
accurate than for the Sun itself.  Neon and argon abundances are taken
from the references given in Pottasch and Bernard-Salas (\cite{pbs}).

\begin{table}[htbp]
\caption[]{Comparison of abundances in \object{NGC\,2392}.}
\begin{tabular}{lrrrr}
\hline
\hline
Elem.  & Present & H$^{\dagger}$  & B$^{\dagger}$ &  Solar$^{\dagger}$  \\ 
\hline  

He & 0.080  & 0.076 & 0.097 &   0.098   \\
C(-4)  & 3.3 & 2.2 & 0.42  &   2.5  \\
N(-4)  & 1.85  & 1.1  & 1.1 &  0.84 \\
O(-4)  & 2.9 & 2.8 & 3.4    &  4.6  \\
Ne(-5) & 8.5 & 6.4 & 7.6  & 12   \\
S(-6)  & 5.0 &     & 4.3  &  14   \\   
Ar(-6) & 2.2 &     & 1.4 &   4.2 \\
Cl(-7) & 1.3 &     &     & 3.5   \\
Si(-6) & 19  &     &     & 32    \\
Fe(-7) & 8.0 &     &     &  280  \\
P(-8)  & 6.5 &     &     &  23   \\

\hline  

\end{tabular}

$^{\dagger}$References: H: Henry et al. (\cite{henry}), B: 
Barker (\cite{barker}),  Solar: Asplund et al. (\cite{asplund}), except Ne and 
Ar (see Pottasch \& Bernard-Salas\,\cite{pbs}).

\end{table}

The main differences with earlier work come about in the
interpretation of the spectrum. First of all because the gradient of
electron temperature has been included here. Considering the fact that
a very large temperature gradient is present the abundances are not
much affected by it. This is probably because the important ions
contributing to the total abundances are formed at temperatures not so
far from the temperature at which the O$^{++}$ ion is formed. An
exception to this is the C$^{++}$ which is formed at a lower
temperature, which explains why Barker (\cite{barker}) has obtained a
much smaller carbon abundance.

A further cause of error is the ICF, can give important errors for Si,
Fe and especially P for which only one ion has been observed. For
these elements the ICF has been determined by comparison with the
results of model calculations made for other PNe which have similar
excitation. This makes the results for these elements uncertain, but
since the ions observed are important contributers to the total
abundance, the error even for these elements is probably less than a
factor of two.

The abundances found agree with those expected from the gradient of PNe
abundances found by Pottasch \& Bernard-Salas\,(\cite{pbs}).

\subsection{Recombination line abundances of carbon}

The C$^{++}$ population can be obtained from the recombination line
$\lambda$4267\AA~as well as from the collisionally excited line at
$\lambda$1909\AA. The advantage of the recombination line is that it
is not sensitive to the electron temperature, and is in the visual
spectrum as well. It has the disadvantage that it is quite faint and
thus difficult to measure accurately. It has been used for at least 30
years, and has been found to sometimes give higher C$^{++}$
populations than the collisional line. The reason for this is not
clear (e.g. see the discussion of Liu et al. (\cite{liu1})). We have
redetermined the C$^{++}$ abundance using using the recombination
rates for this line and the result is shown in the last line of Table
6.  This is essentially the same value as that determined from the
$\lambda$1909\AA~ collisional line, since the error of measurement of
the weak $\lambda$4267\AA~ line is quite large. It is therefore true
that for carbon both determinations give a similar result. This is in
contrast to M\,1-42 where the C$^{++}$ found from the recombination
line is an order of magnitude higher than that found from the
collisional line (Pottasch et al. \cite{pot07}). Since both nebulae
have almost the same electron density it may be concluded that the
electron density does not contribute to this effect.

\section{Discussion of stellar evolution}

The rather low abundances of helium, oxygen, neon and sulfur are
consistent with the position of the nebula in the galaxy which is
about 10 kpc from the galactic center, assuming that the sun is 8 kpc from the 
galactis center. These elements fit very well on
the abundances as a function of position in the galaxy as given by
Pottasch \& Bernard-Salas (\cite{pbs}).  Especially the low helium
abundance indicates that no helium has been produced which implies that the
second dredge-up and hot-bottom burning have not taken place.  The
carbon abundance is somewhat higher than oxygen suggesting that the
third dredge-up has begun. Following the models of Karakas
(\cite{karakas}) this will occur at a stellar mass of about
1.7M\smallsun. This model has increased its nitrogen abundance by
about a factor of five in the first dredge-up so that it is reasonable
that the actual model is to be sought in this direction.

\section{The central star}
\subsection{Stellar temperature}

As discussed in the introduction, the spectrum of the central star has
been studied by several authors. Pauldrach et al. (\cite{paul}) have
fitted model atmospheres to the measured ultraviolet stellar spectrum
and Kudritzki et al. (\cite{kud}) have studied the optical spectrum.
Pauldrach et al.(\cite{paul}) have determined the effective stellar 
temperature from the FeIV/FeV ionization
balance to be 40\,000 K while Kudritzki et al. (\cite{kud}) using the
ionization equilibrium of HeI and HeII obtain a similar value of
45\,000 K.

The stellar temperature can also be determined from the nebular
spectrum.  Enough information is available to compute both the Zanstra
temperature and the Energy Balance temperature of the central star.
The Zanstra temperature requires the knowledge of the stellar apparent
magnitude, the extinction and the H$\beta$ flux. The last two
quantities have already been given in Sect.\,3. The apparent magnitude
is listed by Acker et al. (\cite{acker}) as V$=$10.53. Assuming that
the star radiates as a blackbody the hydrogen Zanstra temperature
$T_{\mathrm{z}}$(H)$=$37\,000 K and the ionized helium Zanstra
temperature is 78\,000 K. The Energy Balance temperature requires the
knowledge of the ratio of the forbidden-line intensities to H$\beta$.
This value is found by summing the intensities given above, and is
about 28 to 30 after making a correction for unmeasured lines using
the table of Preite-Martinez \& Pottasch (\cite{pmp}). This could be
slightly higher but not lower. To convert this value to a stellar
temperature, the formulation of Preite-Martinez \& Pottasch
(\cite{pmp}) is used, assuming blackbody radiation from the central
star. The value of Case II (the nebula is optically thin for radiation
which will ionize hydrogen but optically thick for ionized helium
radiation) for the energy balance temperature ($T_{\mathrm{EB}}$) is
80\,000 K. If a model atmosphere had been used instead of a blackbody,
the energy balance temperature could be lower.  The low value of the
hydrogen Zanstra temperature ($T_{\mathrm{z}}$(H)) may be due to the
nebula being optically thin to radiation ionizing hydrogen. An average
stellar temperature of about 75,000 to 80\,000 K is a reasonable first
approximation. This is also the conclusion of Tinkler \& Lamers
(\cite{tinkler}) who estimate a central star temperature of 74\,000 K.
 
This strange behavior of the central star has been known for many
years.  Heap (\cite{heap}) estimated the stellar spectrum at spectral
class O6 giving the star a much lower temperature than the HeII
Zanstra temperature. This was confirmed by Pottasch et al.
(\cite{pott5}) who pointed out that the stellar continuum between
1500\,\AA~ and 5500\,\AA~ has a spectral distribution more consistent
with a blackbody at 40,000 K than one at 70,000 K. But the measured nebular
spectrum showing ions of high ionization potential
such as \ion{O}{iv} and \ion{Ne}{v} indicates that there is much more
far ultraviolet radiation than is consistent with a blackbody at
40,000 K, a conclusion already reached by Natta et al. (\cite{natta}).

Heap (\cite{heap}) suggested that there is a second star of a higher
temperature which is the source of the nebular ionization. Ciardullo
et al. (\cite{ciar}) using the HST wide field camera have found a
faint companion at a separation of 2.65\arcsec~ in the I band but it
was invisible in the V band. Its properties are as yet unknown. It is
also unknown if it is physically associated with the nebula. It is
therefore possible that the bright star at the center of NGC\,2392 is
not the actual exciting star of the nebula. A determination of the
radius and luminosity of such a star may be derived in the following
Pottasch \& Acker (\cite{pa}). The luminosity of a blackbody of
temperature T=80\,000 K embedded in a nebula is 105 times the
luminosity of the H$\beta$ flux emitted by the nebula.  The luminosity
of the H$\beta$ flux emitted from NGC\,2392, if placed at a distance
of 1500 pc is 4.7L\smallsun. This leads to a total luminosity of the
star L=493L\smallsun. The star must then have a radius of
R=0.116R\smallsun. Such a star would have an unreddened magnitude
m(V)= 14.5. The star found by Ciardullo et al. (\cite{ciar}) was
fainter than m(V)=17.3 so that this could not be the exciting star. If
such an exciting star exists it must be much closer to the bright star
which would make it impossible to observe with the current facilities.

\section{Conclusions}

The nebular abundances of eleven elements have been determined. It has
been shown that those elements which have not been created in stellar
evolution (O, Ne, S, Ar, Cl and possibly Si) have abundances of about
a factor two lower than the solar abundance. This is in line with the
expectation of a gradient in the stellar abundances as a function of
distance from the galactic center.  The helium abundance is lower than
solar; this indicates that no helium has been formed during the
evolution of the central star. Carbon has apparently been formed in a
third dredge-up as it is more abundant than oxygen. Nitrogen has also
been formed, probably in the first dredge-up. The abundances are
consistent with the central star having evolved from an object having
about 1.7M\smallsun. The low iron abundance is also found in almost
all PNe studied and seems to be caused by its being tied up in dust.

The electron temperature is found to vary strongly with the ionization
potential of the ion used for its determination. This indicates a
strong gradient in the nebula, higher than in any PNe yet studied. The
reason for this high gradient is not clear. It could be related to the
fact that diffuse X-ray emission has been detected within its inner
shell (Guerrero et al. \cite{guer}), which in turn may be caused by
the existence of the exceptionally large expansion velocity of the
inner shell.

Finally, the presence of high stages of ionization is confirmed. Both
\ion{O}{iv} and \ion{Ne}{v} are clearly present. This seems to
indicate that the stellar temperature is considerably higher that the
value of 40,000 K found from a study of the central star. The
suggestion that an additional star is present is discussed, but no
solution to this problem can be given.

\section{Acknowledgement}

This work is based on observations made with the Spitzer Space
Telescope, which is operated by the Jet Propulsion Laboratory,
California Institute of Technology under NASA contract 1407. Support
for this work was provided by NASA through Contract Number 1257184
issued by JPL/Caltech.


\begin{thebibliography}{}

\bibitem[1992]{acker}
Acker, A., Marcout, J., Ochsenbein, F. et al. 1992, Strasbourg-ESO catalogue
\bibitem[1979]{aller}
Aller, L.H., \& Czyzak, S.J. 1979, Ap\&SS 62, 397
\bibitem[2005]{asplund}
Asplund, M., Grevesse, N., \& Sauval, A.J. 2005, ASP Conf.\,Ser. (Bash \& Barnes eds.)
\bibitem[1991]{barker}
Barker, T. 1991, ApJ 371, 217
\bibitem[1991]{becker}
Becker, R.H., White, R.L., \& Edwards, A.L. 1991, ApJS 75, 1
\bibitem[1999]{benjamin}
Benjamin, R.A., Skillman, E.D., \& Smits, D.P. 1999, ApJ 514, 307
\bibitem[2001]{bernard}
Bernard Salas, J., Pottasch, S.R., Beintema, D.A., \& Wesselius, P.R. 2001, A\&A 367, 949
\bibitem[1992]{cahn}
Cahn, J.H., Kaler, J.B., \& Stanghellini, L. 1992, A\&AS 94, 399
\bibitem[1999]{ciar}
Ciardullo, R., Bond, H.E., Sipior, M.S. et al. 1999, AJ 118, 488 
\bibitem[1998]{condon}
Condon, J.J., \& Kaplan, D.L. 1998, ApJS 117, 361
\bibitem[2000]{davey}
Davey, A.R., Storey, P.J., \& Kisielius, R. 2000, A\&AS 142, 85
\bibitem[1994]{fluks}
Fluks, M.A., Plez, B., de Winter, D., et al. 1994, A\&AS 105, 311
\bibitem[1991]{greg}
Gregory, P.C., \& Condon, J.J. 1991, ApJS 75, 1011
\bibitem[2005]{guer}
Guerrero, M.A., Chu, Y.-H., Gruendl, R.A., et al. 2005, A\&A 430, L69
\bibitem[1977]{heap}
Heap, S.R., 1977, ApJ 215, 864
\bibitem[2000]{henry}
Henry, R.B.C., Kwitter, K.B., \& Bates, J.A. 2000, ApJ 531, 928
\bibitem[2004]{higdon}
Higdon, S.J.U., Devost, D., Higdon, J.L., et al. 2004, PASP 116, 975
\bibitem[2004]{houck}
Houck, J.R., Appelton, P.N., Armus, L., et al. 2004, ApJS 154, 18
\bibitem[1987]{hummer}
Hummer, D.G., \& Storey, P.J. 1987, MNRAS 224, 801
\bibitem[2003]{karakas}
Karakas, A.I. 2003, Thesis, Monash Univ. Melbourne
\bibitem[2003]{kerber}
Kerber, F., Mignani, R.P., Guglielmetti, F. et al. 2003, A\&A 408, 1029
\bibitem[1994]{kings}
Kingsburgh, R.L., \& Barlow, M.J. 1994, MNRAS 271, 257
\bibitem[1997]{kud}
Kudritzki, R.-P., Mendez, R.H., Puls, J., \& McCarthy, J.K. 1997, IAU\,Symp.180, 64
\bibitem[2000]{liu1}
Liu, X.-W., Storey, P.J., Barlow, M.J. et al. 2000, MNRAS 312, 583
\bibitem[1980]{natta}
Natta, A., Pottasch, S.R., \& Preite-Martinez, A. 1980, A\&A 84, 284
\bibitem[1985]{odell}
O'Dell, C.R., \& Ball, M.E. 1985, ApJ 289, 526
\bibitem[1984]{odellc}
O'Dell, C.R., \& Castaneda, H.O. 1984, ApJ 283, 158
\bibitem[2004]{paul}
Pauldrach, A.W.A., Hoffmann, T.L., \& Mendez, R.H. 2004, A\&A 419, 1111 
\bibitem[1984]{potta}
Pottasch, S.R. 1984, Planetary Nebulae, Reidel Publ. Co. (Dordrecht)
\bibitem[1989]{pa}
Pottasch, S.R., \& Acker, A. 1989 A\&A, 221, 123 
\bibitem[1978]{pott5}
Pottasch, S.R., Wesselius, P.R., Wu, C.C., et al. 1978, A\&A 62, 95
\bibitem[1999]{pott1}
Pottasch, S.R., \& Beintema, D.A. 1999, A\&A 347, 974
\bibitem[2000]{pott2}
Pottasch, S.R., Beintema, D.A., \& Feibelman, W.A. 2000, A\&A 363, 767
\bibitem[2001]{pott4}
Pottasch, S.R., Beintema, D.A., Bernard Salas, J., \& Feibelman,
W.A. 2001, A\&A 380, 684
\bibitem[2006]{pbs}
Pottasch, S.R., \& Bernard-Salas, J. 2006, A\&A, 457, 189
\bibitem[2006]{pot07}
Pottasch, S.R., Bernard-Salas, J., Roellig, T.L. 2006, A\&A, 471, 865
\bibitem[1983]{pmp}
Preite-Martinez, A., \& Pottasch, S.R. 1983 A\&A, 126, 31
\bibitem[1979]{seaton}
Seaton, M.J. 1979, MNRAS 187, 73
\bibitem[2002]{tinkler}
Tinkler, C.M., \& Lamers, H.J.G.L.M. 2002, A\&A 384, 987
\bibitem[2004]{werner}
Werner, M., Roellig, T.L., Low, F.J., et al. 2004, ApJS 154, 1


\end{thebibliography}
\end{document}